\documentclass[twocolumn,amsmath,aps,fleqn]{revtex4}
\usepackage{graphicx,amssymb}
\begin{document}
\newcommand{\be}{\begin{equation}}
\newcommand{\ee}{\end{equation}}
\newcommand{\bq}{\begin{eqnarray}}
\newcommand{\eq}{\end{eqnarray}}
\newcommand{\bsq}{\begin{subequations}}
\newcommand{\esq}{\end{subequations}}
\newcommand{\bc}{\begin{center}}
\newcommand{\ec}{\end{center}}
\newcommand {\R}{{\mathcal R}}
\newcommand{\al}{\alpha}
\newcommand\lsim{\mathrel{\rlap{\lower4pt\hbox{\hskip1pt$\sim$}}
    \raise1pt\hbox{$<$}}}
\newcommand\gsim{\mathrel{\rlap{\lower4pt\hbox{\hskip1pt$\sim$}}
    \raise1pt\hbox{$>$}}}
\newcommand{\newg}{{\tilde g}^{ij}}
\newcommand{\detg}{\sqrt{-{\tilde g}}}
\newcommand{\deterg}{\sqrt{-{g}}}
\newcommand{\lagm}{\mathcal L_m}
\newcommand{\ld}{\lambda}
    
\title{Matter Lagrangian of particles and fluids}

\author{P.P. Avelino}
\email[Electronic address: ]{pedro.avelino@astro.up.pt}
\affiliation{Instituto de Astrof\'{\i}sica e Ci\^encias do Espa{\c c}o, Universidade do Porto, CAUP, Rua das Estrelas, PT4150-762 Porto, Portugal}
\affiliation{Centro de Astrof\'{\i}sica da Universidade do Porto, Rua das Estrelas, PT4150-762 Porto, Portugal}
\affiliation{Departamento de F\'{\i}sica e Astronomia, Faculdade de Ci\^encias, Universidade do Porto, Rua do Campo Alegre 687, PT4169-007 Porto, Portugal}

\author{L. Sousa}
\email[Electronic address: ]{Lara.Sousa@astro.up.pt}
\affiliation{Instituto de Astrof\'{\i}sica e Ci\^encias do Espa{\c c}o, Universidade do Porto, CAUP, Rua das Estrelas, PT4150-762 Porto, Portugal}
\affiliation{Centro de Astrof\'{\i}sica da Universidade do Porto, Rua das Estrelas, PT4150-762 Porto, Portugal}

\date{\today}
\begin{abstract}

We consider a model where particles are described as localized concentrations of energy, with fixed rest mass and structure, which are not significantly affected by their self-induced gravitational field. We show that the volume average of the on-shell matter Lagrangian ${\mathcal L_m}$ describing such particles, in the proper frame, is equal to the volume average of the trace $T$ of the energy-momentum tensor in the same frame, independently of the particle's  structure and constitution. Since both ${\mathcal L_m}$ and $T$ are scalars, and thus independent of the reference frame, this result is also applicable to collections of moving particles and, in particular, to those which can be described by a perfect fluid. Our results are expected to be particularly relevant in the case of modified theories of gravity with nonminimal coupling to matter where the matter Lagrangian appears explicitly in the equations of motion of the gravitational and matter fields, such as $f(R,{\mathcal L_m})$ and $f(R,T)$ gravity. In particular, they indicate that, in this context, $f(R,{\mathcal L_m})$ theories may be regarded as a subclass of $f(R,T)$ gravity.

\end{abstract}
\maketitle 

\section{\label{intr}Introduction}

Precise cosmological observations gathered in recent years have provided us with an increasingly detailed picture of the Universe and its constituents (see, e.g., \cite{Suzuki:2011hu,Anderson:2012sa,Ade:2015xua}). At the present time, the Universe appears to be dominated by two main energy components whose fundamental nature remains mysterious: dark energy (or modified gravity) --- responsible for the current acceleration of the expansion of the Universe --- and dark matter --- required to explain the observed large-scale structure of the Universe. 

However, several other particles, such as baryons and photons, have a much more familiar nature and play a fundamental role in the Universe's structure and evolution. Some of these particles may be regarded as localized energy concentrations, with fixed rest mass and structure, which are not significantly affected by their self-induced gravitational field. Hence, they are often modeled as topological solitons. Still, the modeling of particles as solitons in the simplest scalar field models is not without problems. In particular, the existence of stable finite energy solutions of the nonlinear Klein-Gordon equation in more than one spatial was discarded by Hobard and Derrick \cite{1963PPS....82..201H,1964JMP.....5.1252D} using a simple scaling argument. In Ref. \cite{Avelino:2010bu} Derrick's argument was applied to the case of more general scalar field models and the existence of static global defect solutions of arbitrary dimensionality whose energy does not diverge at spatial infinity was explicitly demonstrated in that context. Skyrmions \cite{1962NucPh..31..556S,Battye:1997qq} --- topological solitons of a Lagrangian embodying chiral symmetry --- and Q-balls \cite{Coleman:1985ki,Kusenko:1997si} --- stationary non-topological solitons whose stability is guaranteed by a conserved charge --- are other examples of localized defects in 3 + 1 dimensions.

In the present paper we start by investigating the necessary conditions for the existence of localized static concentrations  of energy (static solitons) in the absence of a significant self-induced gravitational field, providing a considerable extension of the results presented in Ref. \cite{Avelino:2010bu}. The focus will be on the restrictions imposed on the on-shell matter Lagrangian of a solitonic particle or of a collection of moving solitonic particles which can be described as a fluid. This is particularly relevant for modified theories of gravity with nonminimal coupling to matter where the matter Lagrangian appears explicitly in the equations of motion of the gravitational field, such as $f(R,{\mathcal L_m})$ \cite{Harko:2010mv} and $f(R,T)$ \cite{2011PhRvD..84b4020H} theories of gravity, since, in this context, the knowledge of the energy-momentum tensor is, in general, insufficient to compute the relevant physics \cite{Faraoni:2009rk,Minazzoli:2012md}.

Throughout the paper, we will assume the metric signature $[-,+,\cdots,+]$ and units in which the speed of light in vacuum $c$ equals unity. The Einstein summation convention will be used when a latin or greek index variable appears twice in a single term, once in an upper (superscript) and once in a lower (subscript) position --- the exception will be the latin index $l$ (or $\hat l$), for which the Einstein summation convention shall not be used. Greek and latin indices take the values $\mu, \nu = 0,\cdots,D$; $a, b, c = 1,\cdots,{\mathcal D}$; $i, j, l= 1,\cdots,D$, ${\hat i}, {\hat j}, {\hat l} = N-D+1,\cdots,N$ (with $D \le N$) --- the exception will be the greek index $\lambda$ which shall denote a positive real parameter.

\section{Derrick's argument\label{sec2}}

Consider a $D+1$-dimensional Minkowski space-time with line element given by
\be
ds^2=g_{\mu \nu} dx^\mu dx^\nu =-dt^2 + \delta_{ij} dx^i dx^j
\ee
and a $\mathcal D$-dimensional real scalar field multiplet $\{\phi^1, ..., \phi^{\mathcal D}\}$ 
described by the action $S=\int {\mathcal L}_m \, d^{D+1}x$, where
\be
\label{standardL}
{\mathcal L}_m =X - V(\phi^a)
\ee
is the matter Lagrangian.
Here, $X=-\delta_{ab} \phi^a_{,\mu} \phi^{b,\mu}/2$, the comma in $\phi^a_{,\mu}$ a denotes a partial derivative with respect to the space-time coordinate $x^\mu$, $\phi^a_{,\mu} = g_{\mu\nu} \phi^{a,\nu}$, $g_{\mu\nu}$ are the components of the metric tensor, $\delta_{ab}$ is the Kronecker delta ($\delta_{ab}=1$ if $a=b$ and $\delta_{ab}=0$ if $a \neq b$),  and $V \ge 0$. The energy-momentum tensor for this model is given by
\be
T_{\mu\nu}=\delta_{ab} \phi^a_{,\mu} \phi^b_{,\nu} + g_{\mu\nu}{\mathcal L}_m\,,
\ee
and the total energy can be computed as $E=\int d^D x \, T_{00}$. Consider a static solution $\phi^a=\phi^a(x^i)$ with finite energy equal to
\be
E= \int d^D x \, \left(\delta_{ab} X^{ab} + V (\phi^a) \right) = K + U\,,
\ee
where
\be
K =   \int d^D x \, \left(\delta_{ab} X^{ab} \right)\,, \qquad  U = \int d^D x \, V (\phi^a)  
\ee
are, respectively, the gradient and potential contributions to the total energy, and $X^{ab}=-\phi^a_{,i} \phi^{b,i}/2$.  Under the rescaling $x^i \to {\tilde x}^i=\lambda x^i$, where $\lambda$ is a positive real parameter (that equals unity in the initial configuration), the total energy becomes
\be
E (\lambda)= \int d^D x \, \left(\delta_{ab} X^{ab}_\lambda + V (\phi^a_\lambda) \right)  \,,
\ee
where $\phi^a_\lambda = \phi^a (\lambda x^i)$ and $X^{ab}_\lambda=-\phi^a_{\lambda,i} \phi^{b,i}_\lambda/2$. Changing the integration variable to ${\tilde x}^i=\lambda x^i$, one obtains
\bq
E(\lambda) &=&  \lambda^ {-D} \int d^D {\tilde x} \,  \left(\delta_{ab} \lambda^2 X^{ab} + V (\phi^a) \right) \nonumber  \\
&=&  \lambda^{2-D} K + \lambda^{-D} U   \,.
\eq
A static solution  $\phi^a=\phi^a(x^i)$ must satisfy 
\be\label{Dcondition}
\left[\frac{dE}{d\lambda}\right]_{\lambda=1}=(2-D) K - D U=0\,.
\ee
Hence, no equilibrium static solutions with finite $K > 0$ and finite $U > 0$ exist for $D \ge 2$ \cite{1963PPS....82..201H,1964JMP.....5.1252D}. Despite this fact, static global string and monopole solutions do exist in 3+1 dimensions,  since these are cases for which the gradient energy $K$ formally diverges. Still, in practice, there will always be a cutoff at some energy scale (for instance, in the cosmological context, the linear divergence in the energy of a global monopole has a cutoff due to the finite --- sub-horizon ---  characteristic length of the global monopole network \cite{Lopez-Eiguren:2016jsy,Sousa:2017wvx}).

In Ref. \cite{Avelino:2010bu} Derrick's argument has been generalized to the case of scalar field Lagrangians of the form
\be
\label{gen}
{\mathcal L}_m = {\mathcal L}_m (\phi^a,X^{bc})\,,
\ee
with the energy-momentum tensor given by
\be
T_{\mu\nu}={\mathcal L}_{m,X^{ab}}  \phi^a_{,\mu} \phi^b_{,\nu} + g_{\mu\nu}{\mathcal L}_m\,.
\ee
There, it has been shown that any static equilibrium solution  $\phi^a=\phi^a(x^i)$ must satisfy 
\be\label{Dcondition}
\left[\frac{dE}{d\lambda}\right]_{\lambda=1} = \int d^D x {T^i}_i=0
\ee
or, equivalently, that the average pressure (over volume and directions) must vanish.

\section{Solitonic particles and fluids: $ \langle {\mathcal L_m}\rangle =\langle T \rangle$ \label{sec3}}

Let us describe a static particle as a localized static concentration of energy (static soliton of finite size), and assume that the spacetime is locally Minkowskian on the particle's characteristic lengthscale. Again, we shall implicitly assume that the gravitational field has a negligible impact on the particle structure, so that one may safely neglect the perturbations to the Minkowski metric when computing the total energy of the particle. We shall also start by assuming that the matter fields can be described by a generic real scalar field multiplet $\{\phi^1, ..., \phi^{\mathcal D}\}$, an assumption that shall be relaxed later on.

\subsection{Spherical deformation}

Consider again the transformation $x^i \to {\tilde x}^i = \lambda x^i$, and assume that the matter scalar fields describing a solitonic particle transform under it [this is equivalent to assuming that the functions $\phi^a ({\tilde x}^i)$ are independent of $\lambda$]. The line element may be rewritten as a function of the spatial coordinates ${\tilde x}^i$ as 
\be
ds^2=-dt^2 +  \delta_{ij} dx^i dx^j = -dt^2 +{\tilde g}_{ij} d{\tilde x}^i d{\tilde x}^j\,,
\label{newds}
\ee
where ${\tilde g}_{ij}=\lambda^{-2} \delta_{ij}$. 

Here, we shall also assume that the on-shell matter Lagrangian is invariant with respect to an arbitrary rescaling of the time coordinate, so that
\be
\frac{\delta \mathcal L_m}{\delta g^{0 0}}=0\,, \label{condt}
\ee
in the proper frame in which the particle is static. The components of the energy-momentum tensor of the matter fields are defined by
\be
T_{\mu\nu}=-\frac{2}{\sqrt {-g}} \frac{\delta (\mathcal L_m \sqrt {-g})}{\delta g^{\mu \nu}} = -2 \frac{\delta \mathcal L_m}{\delta g^{\mu \nu}} + g_{\mu \nu} \mathcal L_m\,, \label{Tmunu}
\ee
where $g$ is the determinant of the metric. Equations (\ref{condt}) and (\ref{Tmunu}) imply that the energy density is given by
\be 
\rho=T_{00}=-\mathcal L_m\,, \label{rho}
\ee
so that the total energy of the transformed static concentration of energy is 
\be
E(\lambda)=-\int \mathcal L_m ({{\tilde g}}^{ij},{\tilde x}^k) \detg d^D {\tilde x}\,, \label{energy}
\ee
where $\detg=\lambda^{-D}$. Note that the transformed matter Lagrangian $\mathcal L_m$ will be a function of both ${\tilde g}_{ij}$  and  the matter fields, with the matter fields preserving the dependence on ${\tilde x}^i$ of the initial static configuration.

A necessary condition for static equilibrium around the initial configuration is that $E(\ld)$ has a minimum at $\lambda=1$. Therefore, 
\be
\left[\frac{dE}{d\lambda}\right]_{\lambda=1}=0
\ee
or, equivalently, 
\bq
\left[\frac{dE}{d\lambda}\right]_{\lambda=1} & = & - \int \left[\frac{\partial\left(\lagm \detg\right)}{\partial \ld}\right]_{\ld=1} d^D {\tilde x} = \nonumber \\
& = & -\int \left[\left(\frac{\delta\lagm}{\delta \newg}\frac{\partial \newg}{\partial\ld}-\frac{D}{\ld}\lagm\right)\ld^{-D}\right]_{\ld=1}d^D {\tilde x}=\nonumber\\
& = & -\int \left[2\frac{\delta\lagm}{\delta\newg}\newg-D\lagm\right]d^D {\tilde x}=0\,, \label{DEDL}
\eq
where the fact that ${\tilde g}_{ij}=\lambda^{-2} \delta_{ij}$ (implying that $\partial\newg/\partial\ld=2\newg/\ld$) has been used in the derivation of Eq. (\ref{DEDL}). Hence,
\be
 \int {T^{i}}_{i} d^D x= 0\,, \label{strace}
\ee
which, combined with Eq. (\ref{rho}), implies that
\be
\langle \mathcal L_m \rangle \equiv   \frac{\int \mathcal L_m d^D x}{\int  d^D x}= \frac{\int T  d^D x} {\int  d^D x} \equiv \langle T \rangle\,, \label{trace}
\ee
where $T={T^\mu}_\mu={T^0}_0+{T^i}_i$ is the trace of the energy-momentum tensor. Equation (\ref{trace}) is a scalar equation (${\mathcal L_m}$ and $T$ are both scalars) and, despite being derived in the particle's rest frame, it is also valid in any moving frame. As a matter of fact, since an inertial comoving frame wherein the particle is static exists, the volume averages of ${\mathcal L_m}$ and $T$ are invariant under any Lorentz boost and, thus, Eq.(\ref{trace}) is independent of the velocity of the particle. Therefore, this result is also applicable to fluids that may be well described by a collection of moving solitonic particles, provided that the spacetime is locally Minkowskian on the smallest proper  macroscopic lengthscale in which the fluid approximation applies. Note that here we do not consider potential model-dependent inter-soliton interactions. These, however, are not expected to affect our results unless they have a significant long-range impact on the mass and structure of the particles.

Furthermore, an additional requirement to ensure the stability of the static configuration is that
\be
\left[\frac{d^2E}{d\lambda^2}\right]_{\lambda=1}>0\,,
\ee
which results in the following condition
\bq
\int \left[4\frac{\delta^2\lagm}{\delta(g^{ij})^2}(g^{ij})^2+D\left(D+1\right)\lagm - \right.  \nonumber\\
\left. -\left(4D-2\right)\frac{\delta\lagm}{\delta g^{ij}}g^{ij} \right] d^D x <0\,. \label{lcond}
\eq

The results obtained in this section also hold if the matter fields providing a significant contribution to the energy of the particle include higher order tensor fields ${\bf {\mathcal T}}$ of arbitrary order ${\mathcal N}$, provided that Eq. (\ref{condt}) is satisfied. If, under the transformation $x^i \to {\tilde x}^i = \lambda x^i$, the components ${\mathcal T}_{\mu_1,...,\mu_{\mathcal N}}({\tilde x}^i)$ are assumed to be fixed functions of ${\tilde x}^i$, independently of the value of $\lambda$, then all the results, given by Eqs. (\ref{energy})-(\ref{lcond}), remain valid. 

\subsection{Nonspherical deformation}

Let us now consider the transformation $x^l \to {\tilde x}^l = \lambda_l x^l$, for $l=1,...,D$, where $\lambda_l$ are positive real parameters, such that $\lambda_l=1$ in the initial configuration. The line element, when written as a function ${\tilde x}^i$, is still given by Eq. (\ref{newds}), but in this case, $\newg= \ld_i\ld_j \delta^{ij}$. We shall demonstrate in the present  section that considering this more general deformation, allowing for different directional scaling parameters, leads to conditions on the form of the energy-momentum tensor that are even more restrictive than those in Eq.  (\ref{strace}).

Assuming that Eq. (\ref{condt}) remains valid, the total energy of the transformed static configuration may be written as
\be
E\left(\ld_1,...,\ld_D\right)=-\int \lagm\left(\newg,{\tilde x}\right)\detg d^D{\tilde x}\,,
\ee
with
\be
\detg = \prod _{i}\ld_i^{-1}\,.
\ee
In this case, static equilibrium can only be preserved if
\be
\left[\frac{dE}{d\ld_i}\right]_{\ld_1=1,...,\ld_D=1}=0\,,\qquad \mbox{for all }i=1,...D\,.
\ee

Considering a specific value of $l$ and applying a similar procedure to that employed in Eq. (\ref{DEDL}) one obtains
\be
-\int\left(2\frac{\delta\lagm}{\delta g^{ll}}  g^{ll} -\lagm\right)d^D x=0\,,
\ee
or, equivalently,
\be
\int T_{l l} d^D x =0\,,\qquad \mbox{for all } l=1,...,D\,.
\ee
This not only implies that the volume of the spatial trace of the energy-momentum tensor must be equal to zero in the rest frame of the solitonic particle [Eq. (\ref{strace})] but also that the volume average of the pressure along all $l=1,...,D$ directions must vanish.

Moreover, static equilibrium around the initial static configuration can only be guaranteed if $E(\ld_1,...,\ld_d)$ has a minimum at $\ld_1=...=\ld_D=1$, implying that
\be
\left[\frac{d^2 E}{d\ld_i^2}\right]_{\ld_1=1,...,\ld_D=1}>0\,,
\ee
which results in the following constraints
\be
\int \left[4\frac{\delta^2\lagm}{\delta (g^{ll})^2}(g^{ll})^2+2\lagm-2\frac{\delta\lagm}{\delta g^{ll}}\right]d^D x<0\,,
\ee
for all $l=1,...,D$.

\section{Defects of codimension $D$ in $N+1$-dimensional space-times \label{sec4}}

Our results may be generalized to also describe p-branes of codimension $D$ ($p=N-D$),  embedded in a Minkowski space-time with $N > D$ spatial dimensions (see Refs.  \cite{Sousa:2011ew,Sousa:2011iu} for a unified unified framework describing the macroscopic evolution of featureless p-branes). Assuming that
\be
\frac{\delta \mathcal L_m}{\delta g^{{\hat i} {\hat j} }}=0\,, \qquad \mbox{for all } {\hat i},{\hat j} =D+1,...,N\,, \label{defect}
\ee
$x^{\hat i}$ with ${\hat i}=D+1,...,N$ being the additional space-time coordinates, Eq. (\ref{Tmunu}) implies that
\be
T_{{\hat l}{\hat l}} = \mathcal L_m\,, \qquad \mbox{for all } {\hat l}=D+1,...,N\,,
\ee
independently of the velocity of the observer. In practice Eq. (\ref{defect}) means that the defect is featureless along the ${\hat l}=D+1,...,N$ directions or, equivalently, that it is not possible to measure the velocity of the defect along these directions. In the defect rest frame, one has that $T_{00}= -T_{{\hat l}{\hat l}}$ and $T_{ll}=0$. Hence, a statistically homogeneous and isotropic network of frozen defects will have an (averaged) equation of state given by
\be
p=-\frac{N-D}{N} \rho\,.
\ee
Here, $\rho$ and $p$ represent the average energy density and pressure associated with the defect network,
independently of the specific form of the matter Lagrangian or the defect geometry along the first $D$ spatial directions. If the defects have a nonzero root mean square velocity $v$, the (average) pressure becomes \cite{Avelino:2015kdn}
\be
p=\left(-\frac{N-D}{N} + \frac{N-D+1}{N} v^2\right)\rho \,,
\ee
so that $p \to \rho/N$ in the $v \to 1$ limit (note that if $N=3$ and $v=1$ then $p=\rho/3$).

\section{Conclusions \label{sec5}}

In this paper, we have shown that the volume average of the matter Lagrangian ${\mathcal L_m}$ of a solitonic particle, or of a collection of solitonic particles with fixed rest mass and structure, is equal to the volume average of the trace $T$ of the particle's energy-momentum tensor. This result, obtained with minimal assumptions about the particle structure and constitution, is crucial for the accurate computation of the equations of motion of the gravitational and matter fields in the context of modified theories of gravity with nonminimal coupling to matter where the matter Lagrangian appears explicitly in the equations of motion of the gravitational field, such as $f(R,{\mathcal L_m})$ and $f(R,T)$ gravity. It also implies that, whenever the sole contribution to the gravitational field comes from matter sources which may be well modeled by a collection of solitonic particles with fixed rest mass and structure, $f(R,{\mathcal L_m})$ gravity may be considered a subclass of $f(R,T)$ gravity.\\

P.P.A. thanks Rui Azevedo for enlightening discussions. L.S. was supported by Funda{\c c}\~ao para a Ci\^encia e a Tecnologia (FCT, Portugal) through the Grants No.  SFRH/BPD/76324/2011 and No. CIAAUP-02/2018-PPD. Funding of this work has also been provided by the FCT Grant No. UID/FIS/04434/2013. This paper benefited from the participation of the authors on the European Cooperation in Science and Technology (COST) action CA15117 (CANTATA), supported by COST. 

\bibliography{PL}

\begin{thebibliography}{19}
\expandafter\ifx\csname natexlab\endcsname\relax\def\natexlab#1{#1}\fi
\expandafter\ifx\csname bibnamefont\endcsname\relax
  \def\bibnamefont#1{#1}\fi
\expandafter\ifx\csname bibfnamefont\endcsname\relax
  \def\bibfnamefont#1{#1}\fi
\expandafter\ifx\csname citenamefont\endcsname\relax
  \def\citenamefont#1{#1}\fi
\expandafter\ifx\csname url\endcsname\relax
  \def\url#1{\texttt{#1}}\fi
\expandafter\ifx\csname urlprefix\endcsname\relax\def\urlprefix{URL }\fi
\providecommand{\bibinfo}[2]{#2}
\providecommand{\eprint}[2][]{\url{#2}}

\bibitem[{\citenamefont{Suzuki et~al.}(2012)}]{Suzuki:2011hu}
\bibinfo{author}{\bibfnamefont{N.}~\bibnamefont{Suzuki}} \bibnamefont{et~al.},
  \bibinfo{journal}{Astrophys.J.} \textbf{\bibinfo{volume}{746}},
  \bibinfo{pages}{85} (\bibinfo{year}{2012}).

\bibitem[{\citenamefont{Anderson et~al.}(2013)}]{Anderson:2012sa}
\bibinfo{author}{\bibfnamefont{L.}~\bibnamefont{Anderson}}
  \bibnamefont{et~al.}, \bibinfo{journal}{Mon. Not. R. Astron. Soc.}
  \textbf{\bibinfo{volume}{427}}, \bibinfo{pages}{3435} (\bibinfo{year}{2013}).

\bibitem[{\citenamefont{Ade et~al.}(2016)}]{Ade:2015xua}
\bibinfo{author}{\bibfnamefont{P.~A.~R.} \bibnamefont{Ade}}
  \bibnamefont{et~al.} (\bibinfo{collaboration}{Planck Collaboration}),
  \bibinfo{journal}{Astron. Astrophys.} \textbf{\bibinfo{volume}{594}},
  \bibinfo{pages}{A13} (\bibinfo{year}{2016}).

\bibitem[{\citenamefont{{Hobart}}(1963)}]{1963PPS....82..201H}
\bibinfo{author}{\bibfnamefont{R.~H.} \bibnamefont{{Hobart}}},
  \bibinfo{journal}{Proc. Phys. Soc.} \textbf{\bibinfo{volume}{82}},
  \bibinfo{pages}{201} (\bibinfo{year}{1963}).

\bibitem[{\citenamefont{{Derrick}}(1964)}]{1964JMP.....5.1252D}
\bibinfo{author}{\bibfnamefont{G.~H.} \bibnamefont{{Derrick}}},
  \bibinfo{journal}{J. Math. Phys.} \textbf{\bibinfo{volume}{5}},
  \bibinfo{pages}{1252} (\bibinfo{year}{1964}).

\bibitem[{\citenamefont{Avelino et~al.}(2011)\citenamefont{Avelino, Bazeia, and
  Menezes}}]{Avelino:2010bu}
\bibinfo{author}{\bibfnamefont{P.~P.} \bibnamefont{Avelino}},
  \bibinfo{author}{\bibfnamefont{D.}~\bibnamefont{Bazeia}}, \bibnamefont{and}
  \bibinfo{author}{\bibfnamefont{R.}~\bibnamefont{Menezes}},
  \bibinfo{journal}{Eur. Phys. J. C} \textbf{\bibinfo{volume}{71}},
  \bibinfo{pages}{1683} (\bibinfo{year}{2011}).

\bibitem[{\citenamefont{{Skyrme}}(1962)}]{1962NucPh..31..556S}
\bibinfo{author}{\bibfnamefont{T.~H.~R.} \bibnamefont{{Skyrme}}},
  \bibinfo{journal}{Nucl. Phys.} pp. \bibinfo{pages}{556--569}
  (\bibinfo{year}{1962}).

\bibitem[{\citenamefont{Battye and Sutcliffe}(1997)}]{Battye:1997qq}
\bibinfo{author}{\bibfnamefont{R.~A.} \bibnamefont{Battye}} \bibnamefont{and}
  \bibinfo{author}{\bibfnamefont{P.~M.} \bibnamefont{Sutcliffe}},
  \bibinfo{journal}{Phys. Rev. Lett.} \textbf{\bibinfo{volume}{79}},
  \bibinfo{pages}{363} (\bibinfo{year}{1997}).

\bibitem[{\citenamefont{Coleman}(1986)}]{Coleman:1985ki}
\bibinfo{author}{\bibfnamefont{S.~R.} \bibnamefont{Coleman}},
  \bibinfo{journal}{Nucl. Phys. {\bf B262}, 263 (1985); Nucl. Phys. {\bf B269},
  744 (E)}  (\bibinfo{year}{1986}).

\bibitem[{\citenamefont{Kusenko and Shaposhnikov}(1998)}]{Kusenko:1997si}
\bibinfo{author}{\bibfnamefont{A.}~\bibnamefont{Kusenko}} \bibnamefont{and}
  \bibinfo{author}{\bibfnamefont{M.~E.} \bibnamefont{Shaposhnikov}},
  \bibinfo{journal}{Phys. Lett. B} \textbf{\bibinfo{volume}{418}},
  \bibinfo{pages}{46} (\bibinfo{year}{1998}).

\bibitem[{\citenamefont{Harko and Lobo}(2010)}]{Harko:2010mv}
\bibinfo{author}{\bibfnamefont{T.}~\bibnamefont{Harko}} \bibnamefont{and}
  \bibinfo{author}{\bibfnamefont{F.~S.~N.} \bibnamefont{Lobo}},
  \bibinfo{journal}{Eur. Phys. J. C} \textbf{\bibinfo{volume}{70}},
  \bibinfo{pages}{373} (\bibinfo{year}{2010}).

\bibitem[{\citenamefont{{Harko} et~al.}(2011)\citenamefont{{Harko}, {Lobo},
  {Nojiri}, and {Odintsov}}}]{2011PhRvD..84b4020H}
\bibinfo{author}{\bibfnamefont{T.}~\bibnamefont{{Harko}}},
  \bibinfo{author}{\bibfnamefont{F.~S.~N.} \bibnamefont{{Lobo}}},
  \bibinfo{author}{\bibfnamefont{S.}~\bibnamefont{{Nojiri}}}, \bibnamefont{and}
  \bibinfo{author}{\bibfnamefont{S.~D.} \bibnamefont{{Odintsov}}},
  \bibinfo{journal}{Phys. Rev. D} \textbf{\bibinfo{volume}{84}},
  \bibinfo{eid}{024020} (\bibinfo{year}{2011}).

\bibitem[{\citenamefont{Faraoni}(2009)}]{Faraoni:2009rk}
\bibinfo{author}{\bibfnamefont{V.}~\bibnamefont{Faraoni}},
  \bibinfo{journal}{Phys. Rev. D} \textbf{\bibinfo{volume}{80}},
  \bibinfo{pages}{124040} (\bibinfo{year}{2009}).

\bibitem[{\citenamefont{Minazzoli and Harko}(2012)}]{Minazzoli:2012md}
\bibinfo{author}{\bibfnamefont{O.}~\bibnamefont{Minazzoli}} \bibnamefont{and}
  \bibinfo{author}{\bibfnamefont{T.}~\bibnamefont{Harko}},
  \bibinfo{journal}{Phys. Rev. D} \textbf{\bibinfo{volume}{86}},
  \bibinfo{pages}{087502} (\bibinfo{year}{2012}).

\bibitem[{\citenamefont{Lopez-Eiguren et~al.}(2017)\citenamefont{Lopez-Eiguren,
  Urrestilla, and Achúcarro}}]{Lopez-Eiguren:2016jsy}
\bibinfo{author}{\bibfnamefont{A.}~\bibnamefont{Lopez-Eiguren}},
  \bibinfo{author}{\bibfnamefont{J.}~\bibnamefont{Urrestilla}},
  \bibnamefont{and}
  \bibinfo{author}{\bibfnamefont{A.}~\bibnamefont{Achúcarro}},
  \bibinfo{journal}{J. Cosmol. Astropart. Phys. 01 (2017) 20; J. Cosmol.
  Astropart. Phys. 06 E01}  (\bibinfo{year}{2017}).

\bibitem[{\citenamefont{Sousa and Avelino}(2017)}]{Sousa:2017wvx}
\bibinfo{author}{\bibfnamefont{L.}~\bibnamefont{Sousa}} \bibnamefont{and}
  \bibinfo{author}{\bibfnamefont{P.~P.} \bibnamefont{Avelino}},
  \bibinfo{journal}{Phys. Rev. D} \textbf{\bibinfo{volume}{96}},
  \bibinfo{pages}{023521} (\bibinfo{year}{2017}).

\bibitem[{\citenamefont{Sousa and Avelino}(2011{\natexlab{a}})}]{Sousa:2011ew}
\bibinfo{author}{\bibfnamefont{L.}~\bibnamefont{Sousa}} \bibnamefont{and}
  \bibinfo{author}{\bibfnamefont{P.~P.} \bibnamefont{Avelino}},
  \bibinfo{journal}{Phys. Rev. D} \textbf{\bibinfo{volume}{83}},
  \bibinfo{pages}{103507} (\bibinfo{year}{2011}{\natexlab{a}}).

\bibitem[{\citenamefont{Sousa and Avelino}(2011{\natexlab{b}})}]{Sousa:2011iu}
\bibinfo{author}{\bibfnamefont{L.}~\bibnamefont{Sousa}} \bibnamefont{and}
  \bibinfo{author}{\bibfnamefont{P.~P.} \bibnamefont{Avelino}},
  \bibinfo{journal}{Phys. Rev. D} \textbf{\bibinfo{volume}{84}},
  \bibinfo{pages}{063502} (\bibinfo{year}{2011}{\natexlab{b}}).

\bibitem[{\citenamefont{Avelino and Sousa}(2016)}]{Avelino:2015kdn}
\bibinfo{author}{\bibfnamefont{P.~P.} \bibnamefont{Avelino}} \bibnamefont{and}
  \bibinfo{author}{\bibfnamefont{L.}~\bibnamefont{Sousa}},
  \bibinfo{journal}{Phys. Rev. D} \textbf{\bibinfo{volume}{93}},
  \bibinfo{pages}{023519} (\bibinfo{year}{2016}).

\end{thebibliography}

\end{document}